\documentclass[%
superscriptaddress,
preprint,
amsmath,amssymb,
aps,
prb,
]{revtex4-1}

\usepackage{graphicx}
\usepackage{dcolumn}
\usepackage{bm}

\begin{document}
\title{A simple expansion method for understanding toroidal multipoles and anapoles in light-matter interactions}

\author{Shi-Qiang Li}
\email{shiqiang.li@unimelb.edu.au}
\affiliation{
	Department of Electrical and Electronic Engineering \\
	University of Melbourne, Victoria 3010 \\
	Australia
}%
\author{Kenneth B. Crozier}%
\email{kenneth.crozier@unimelb.edu.au}
\affiliation{
	Department of Electrical and Electronic Engineering \\
	University of Melbourne, Victoria 3010 \\
	Australia
}%
\affiliation{
	School of Physics \\
	University of Melbourne, Victoria 3010 \\
	Australia
}%

\thanks{This work was supported by the Australian Research Council (DP150103736 and FT140100577) and by the Victorian Endowment for Science, Knowledge and Innovation (VESKI). The authors thank Dr. Patrick Grahn. from COMSOL Inc. and Prof. John B. Ketterson from Northwestern University for helpful discussion.}

\date{\today}

\begin{abstract}
Toroidal multipoles are a topic of increasing interest in the nanophotonics and metamaterials communities. In this work, we separate out the toroidal multipole components of multipole expansions in polar coordinates (two- and three-dimensional) by expanding the Bessel or spherical Bessel functions. We discuss the formation of magnetic anapoles from the interaction between the magnetic toroidal dipole and the magnetic dipole.  Our method also reveals that there are higher order current configurations other than the toroidal electric multipole that have the same radiation characteristics as the pure electric dipole.
\end{abstract}

\maketitle


\section{Introduction}

The toroidal multipole is a class of complex current configurations. The fundamental order of the toroidal multipole is a toroidal dipole, whose current flow is similar to current flowing inside a solenoid coil bent into the shape of a torus. The radiation pattern of a toroidal dipole is the same as an electric dipole, and paired strengths of the two result in a total cancellation of radiation. This condition is a non-radiating but non-trivial current configuration that is called an anapole. It was first proposed in atomic physics by Zeldovich \cite{zeldovich58} and observed later in cesium\cite{RN108}. The toroidal multipole has also been found responsible for some unusual phenomena in condensed matter physics \cite{RN106,RN107,RN63,RN65} and is currently the topic of much interest in the nanophotonics and metamaterials communities\cite{kaelberer10, miroshnichenko15, papsimakis16}. 

One of the most prominent characteristics of the toroidal multipoles is that each of the multipoles has the same radiation pattern as its Cartesian electric multipole counterpart, but has a very different current configuration and scales with a different order of $kr$, where $k$ is the wavenumber and $r$ is the radius of the scatterer \cite{fernandez15, papsimakis16, naz16, vrejoiu2002,radescu02}. The toroidal dipole moment can be picked out from the multipole expansion in Cartesian coordinates\cite{RN96, grahn12}. However, the complexity of the extraction makes it difficult to obtain higher order toroidal moments (i.e. beyond dipoles). By contrast, the classical multipole expansion in spherical coordinates was discovered more than one century ago \cite{mie1908}. It expands an arbitrary radiation pattern as an incoherent summation of orthogonal radiation patterns. Therefore, the Cartesian electric multipole and the toroidal electric multipole \cite{RN134}, which have the same radiation pattern, will be contained in the same term of the expansion. 

In this work, we separate out the toroidal multipole in the classical multipole expansion of the scattering current. We expect that our method will be useful for any application that involves toroidal multipoles including magnetoelectricity \cite{RN138, RN139, RN140, RN141}, metamaterials \cite{kaelberer10, papsimakis16,RN142}, and  nanophotonics \cite{RN135,RN137,mirzaei15}. The structure of our article is as follows. We first derive the two-dimensional (2-D) multipole expansion in polar coordinates, which is simple to perform and understand. We then extend the method to three-dimensional (3-D) spherical coordinates. Following this, we present examples of application of this method to 2-D (nanotube with both electric and magnetic anapole excitations) and 3-D (sphere with anapole excitation) cases. 

\section{Multipole expansion in 2-D polar and 3-D spherical coordinates}
\label{sec:multipole-in-2D}

Scattering from objects can be treated as radiation from scattering
currents. In bounded regions with scattering current sources, we only need to consider the source electric current, as long as the relative permeability \(\mu\) is 1, which is
valid for non-magnetic materials. Following Jackson's formulation\cite{jackson99}, we obtain:
\begin{subequations}
	\begin{eqnarray}
	(\nabla^2 +k^2)\vec E' =& -\frac{i}{k}\eta \nabla \times \nabla \times  \vec{\bf J} \label{eq:three},
	\\
	\nabla \cdotp\vec E' =& 0 \label{eq:four},
	\\
	(\nabla^2 +k^2)\vec H =& -\nabla \times  \vec{\bf J} \label{eq:one} , 
	\\
	\nabla \cdotp\vec H =& 0 \label{eq:two}
	\end{eqnarray}
\end{subequations}
, where $k$ is the wavenumber, $\eta$ is the wave impedance $\sqrt{\mu_0/(\epsilon_0\epsilon_h)}$, \(\vec{\bf J}\) is the scattering current, defined as
$\vec{\bf J} = -i\omega \epsilon_0\left(\epsilon _{\vec r} - \epsilon_{h}\right)\vec E$ \cite{grahn12}, and $\vec E' = \vec E + \frac{i}{\omega\epsilon_0 \epsilon_h} \vec {\bf J}$. $\omega$ is the angular frequency of the wave under consideration, $\epsilon _{\vec r}$ is the relative permittivity at the position $\vec{r}$ and $\epsilon _{h}$ is the relative permittivity of the host medium. $\mu_0$ and  $\epsilon_0$ are the permeability and the permittivity of the free space respectively. 
This technique is also referred to as the volume-current method for solving scattering problems\cite{johnson01, snyder70}.   

In 2-D polar coordinates, Eq~\ref{eq:three} to Eq~\ref{eq:two} can be solved separately for the TE wave (electric field $\vec E = E_z \hat z$) and the TM wave ($\vec H = H_z \hat z$) to obtain the following multipole coefficients
\begin{subequations}
	\begin{eqnarray}
	A_m =&\frac{\eta k}{4}\int  {\bf J_z} e^{im\psi}\left[-J_m\right] d\bf s \label{eq:two_a},
	\\
	B_m =& \frac{i}{4}\int  \frac{e^{im\psi}}{r}\left[\left(kr\right) {\bf {J_\psi}} J_{m+1} +m\left(i{\bf J_r}-{\bf {J_\psi}}\right)J_m\right]d\bf s \label{eq:two_b} 
	\\
	E_z =& \sum_{m=-\infty}^{m=+\infty} A_mH_m^{\left(1\right)}e^{im\psi} \label{eq:two_c}
	\\
	H_z =& \sum_{m=-\infty}^{m=+\infty} B_mH_m^{\left(1\right)}e^{im\psi} \label{eq:two_d}
	\end{eqnarray}
\end{subequations}
, where \(\vec r = \sqrt{(x^2 + y^2)} \left(cos(\psi)\hat x + sin(\psi) \hat y\right)\), \(\vec \psi = cos(\psi)\hat x + sin(\psi) \hat y\), $H_m$ is the Hankel function of the first kind with argument $kr$, and  $J_m$ is the Bessel function of the first kind with argument $kr$ and order $m$. We use the boldface letter $\bf J$ to represent the scattering current, with a subscript that indicates which component of the three principal axes of cylindrical coordinates is being referred to. A detailed derivation is presented in Section 1 of the Appendix. With these equations, we can calculate the partial scattering cross-section from each individual multipole $m$ for any current distributions.   

Recall that the Cartesian multipole expansion is expanded in terms of the order of $kr$, while the Mie expansion is performed with respect to orthogonal radiation patterns\cite{grahn12}. Since the Cartesian toroidal multipole has the same radiation pattern as its corresponding electric multipole but scales with two more orders of $kr$\cite{papsimakis16}, a natural way to separate the Cartesian toroidal multipole from the spherical electric multipole is then to group the terms in the integrands on the right-hand sides of Eq~\ref{eq:two_a} and \ref{eq:two_b} by their order of $kr$. However, care must be taken here because the Bessel function is dependent on $kr$, thus it is necessary to expand the Bessel function $J_m$ in powers of $kr$ by the method of Frobenius \cite{RN88},
\begin{eqnarray}
J_{m} = && \frac{1}{m!}\left(\frac{kr}{2}\right)^m - \frac{1}{\left(m+1\right)!}\left(\frac{kr}{2}\right)^{m+2} + ...
\end{eqnarray}
We denote the leading term $\frac{1}{m!}\left(\frac{kr}{2}\right)^m$ as $J_{m,0}$ and the second term $- \frac{1}{\left(m+1\right)!}\left(\frac{kr}{2}\right)^{m+2}$ as $J_{m,1}$. This enables direct comparison with the results of the Cartesian multipole expansion \cite{radescu02, naz16, papsimakis16}, by keeping the two leading terms in the expansion, as shown in Eq~\ref{eq:14a} and Eq~\ref{eq:13a}: 
\begin{subequations}
	\begin{eqnarray}
	A_m \approx&&\frac{\eta k}{4}\int  {\bf J_z}e^{im\psi}\left[-\left(J_{m,0}+\underline{J_{m,1}}\right)\right] d\bf s \label{eq:14a},
	\\
	B_m \approx&& \frac{i}{4}\int  \frac{e^{im\psi}}{r} [\left(kr\right) {\bf J_\psi} \underline{J_{m+1,0}} \nonumber
	\\+ && m\left(i{\bf J_r}-{\bf J_\psi}\right)\left(J_{m,0}+\underline{J_{m,1}}\right)]d\bf s \label{eq:13a}.
	\end{eqnarray}
\end{subequations}
In Eq~\ref{eq:14a} and Eq~\ref{eq:13a}, the underlined terms have a dependence on $kr$, which is two orders higher $kr$ (i.e. by $(kr)^2$) than the terms without underlines. We can now write these coefficients down explicitly for m=0 and m=1, in a manner that separates the pure dipole term from the toroidal term. This is done as Eqs 5a and 5b (for TE mode) and Eqs 5c and 5b (for TM mode). The top and bottom rows of each equation are the pure dipole term and toroidal term, respectively. Note that we need to treat $B_0$ as a special case, as the second term in Eq~\ref{eq:13a} vanishes when $m=0$. We therefore include both $J_{1,0} $ and $J_{1,1}$ in the expansion. We observe that the terms in the bottom rows of Eq~ 5a-d are two orders of $kr$ higher than those in the top rows, and they have the same radiation pattern. These match the characteristics of the known toroidal terms. Based on the order (i.e. $kr$ dependence) and the polarization of the radiated field, we can further identify Eq.~\ref{eq:2DsepA} and Eq.~\ref{eq:2DsepD} as being the electric dipoles/toroidal dipoles and Eq.~\ref{eq:2DsepB} and Eq.~\ref{eq:2DsepC} as being the magnetic dipoles/toroidal dipoles. 
\begin{subequations}
	\begin{eqnarray}
	A_0 =&& \frac{\eta k}{4}\int  \begin{Bmatrix}
	-1 \\
	- \frac{\left(kr\right)^2}{4}
	\end{Bmatrix} {\bf J_z} d\boldsymbol{s} \label{eq:2DsepA}	
	\\
	A_1 =&& \frac{\eta k}{4}\int  \begin{Bmatrix}
	-\frac{kr}{2} \\
	+ \frac{1}{2}\left(\frac{kr}{2}\right)^3
	\end{Bmatrix} {\bf J_z} e^{i\psi} d\boldsymbol{s} \label{eq:2DsepB}	
	\\
	B_0 =&& \frac{ik}{4}\int \begin{Bmatrix}\frac{kr}{2} \\ -\frac{\left(kr\right)^3}{16}\end{Bmatrix} {\bf J_\psi} d\boldsymbol{s}  \label{eq:2DsepC}	
	\\
	B_1 =&& \frac{ik}{4}\int  \begin{Bmatrix}
	\left(i{\bf J_r} - {\bf J_\psi} \right) \\
	- \frac{\left(kr\right)^2}{2} \left(i{\bf J_r} - 3/2 {\bf J_\psi}\right)
	\end{Bmatrix} \frac{e^{i\psi}}{2} d\boldsymbol{s} \label{eq:2DsepD}	
	\end{eqnarray}
\end{subequations}

In general, the separated terms found by the expansion of Bessel functions are not orthogonal. This is different from the original spherical multipole expansion, in which the incoherent summation of the partial scattering cross-sections from different multipoles gives the radiated power. A cross-term of the electric and toroidal multipole may be used to account for the interaction between the terms with the same radiation pattern \cite{radescu02}. That is the origin of the anapole condition\cite{afanasiev95, miroshnichenko15, wang16}, for which there is destructive interference between the electric dipole and toroidal dipole.

Within the same framework, we find that the 3-D spherical expansion can also be expressed in a similar manner as follows (readers interested in the derivation should refer to the Appendix Section 3), 

\begin{subequations}
	\begin{eqnarray}
	a_{E,d}\left(l,m\right) =  \frac{\left(-i\right)^{l-1}k^2\eta O_{lm}}{\left[\pi\left(2l+1\right)\right]^{1/2}} \nonumber 
	\int_{\bf v} e^{im\psi}\frac{\left(l+1\right)}{kr}j_{l,0}\bigg\{lP_l^m\left(\cos \theta \right){\bf J_r} 
	+\tau_{lm}\left(\theta\right){\bf J_\theta}+i\pi_{lm}\left(\theta\right){\bf J_\psi}\bigg\} d {\bf v} \nonumber\\
	\label{eq:16}
	\end{eqnarray}

	\begin{eqnarray}
	a_{E,t}\left(l,m\right) = \frac{\left(-i\right)^{l-1}k^2\eta O_{lm}}{\left[\pi\left(2l+1\right)\right]^{1/2}} 
	\int_{\bf v} e^{im\psi}\frac{-kr}{4l+6}j_{l,0} \bigg\{&& \left(l+1\right) P_l^m\left(\cos \theta \right){\bf J_r} \nonumber\\
	+ && \left(l+3\right)\left[\tau_{lm}\left(\theta\right){\bf J_\theta}+i\pi_{lm}\left(\theta\right){\bf J_\psi}\right]\bigg\} d {\bf v}  
	\label{eq:22}
	\end{eqnarray}
\end{subequations}

, where $P_l^m(\cos \theta)$ are the associated Legendre polynomials \cite{jackson99}. $a_{E,d}\left(l,m\right)$ is the term corresponding to the Cartesian electric multipole and $a_{E,t}\left(l,m\right)$ corresponds to the toroidal multipole. The other three terms in the equation are:  $O_{lm} = \frac{\left(2l+1\right)\left(l-m\right)!}{4\pi \left[l\left(l+1\right)\right]\left(l+m\right)}$, $\tau_{lm} = \frac{d}{d\theta}P_l^m\left(\cos\theta\right)$, and $\pi_{lm} = \frac{m}{\sin\theta}P_l^m\left(\cos\theta\right)$. We denote the small argument limit of the spherical Bessel function as $j_{l,0} =  \frac{ \left(2kr\right)^l l!}{\left(2l+1\right)!}$. 

\section{Examples}
\label{sec:simulations}
We next verify our method by applying the above equations to two example light scattering problems. We use the commercial finite element method solver COMSOL for two purposes. First, the scattering current distribution is found by numerically solving Maxwell's equations. Second, our theory is used to find the partial scattering cross section from the scattering current. A two-dimensional problem and a three-dimensional problem are considered, both involving anapoles.

\begin{figure}
	\centering
	\includegraphics[scale=0.6]{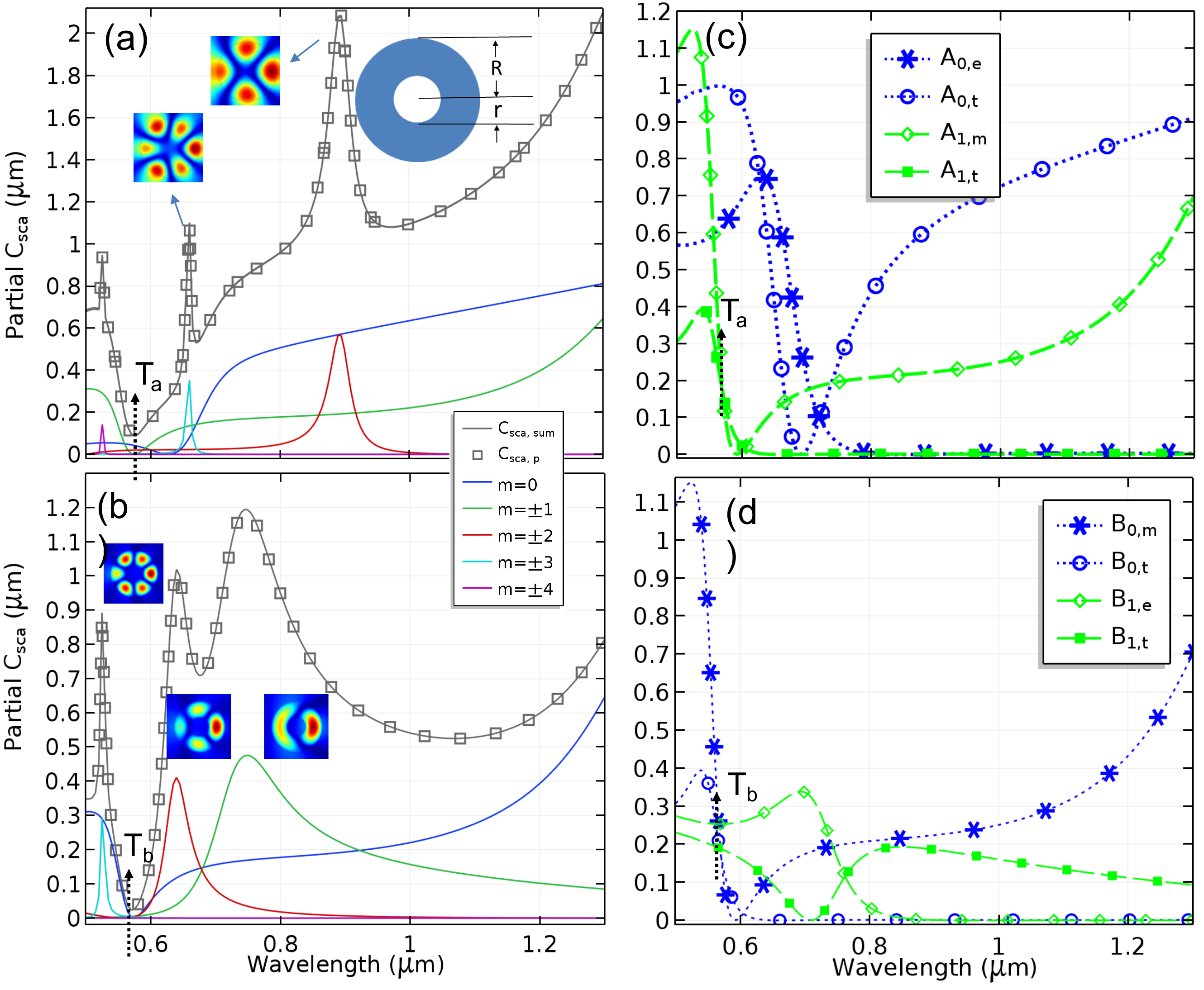}
	
	\caption{ Decomposition of the scattering cross-section of a nanotube in the 2D multipole expansion. The incident plane wave has a wavevector along $\hat x$ direction. TE wave case is shown as panel (a) and TM wave case is shown as panel (b). Inset of (a) shows a schematic of a cross-section of the nanotube, whose inner r and outer R radii are 50 nm and 150 nm, respectively. The host medium is air with n = 1 and the tube has a refractive index of 3.5. Insets in panels (a) and (b) are mode profiles of various orders (amplitude of the electric field), with details explained in the text. The expanded toroidal terms and Cartesian dipole terms are plotted in Figure (c) and (d) for TE and TM cases respectively. These are partial scattering cross-sections calculated based on the separation shown in Eq. 5.}
	\label{fig:2dmultipole}
\end{figure}
To demonstrate the application of our method in two dimensions and the anapole condition, we choose to study a hollow cylinder that we term a nanotube (Fig.~\ref{fig:2dmultipole}). We plot the partial scattering cross-section with $m$ ranging from 0 to 4 (shown as the solid line plots) with $A_m$ and $B_m$ found from Eq~\ref{eq:two_a} and Eq~\ref{eq:two_b}, in Fig.~\ref{fig:2dmultipole} (a) and (b). We also plot the result calculated from the definition of scattering cross-section (gray solid line) , i.e. integrating the Poynting vector of the scattered fields over a contour enclosing the nanotube. It is in complete agreement with the result obtained by summing the multipoles from $m$ = 0 to 4 (shown as the square markers in Fig.~\ref{fig:2dmultipole}). Note that due to the circular symmetry of the nanotube, the positive and negative orders of $m$ are degenerate. For both the TE and the TM polarizations, there are three very distinct peaks and one distinct dip in the scattering spectra. The dips are denoted with black dashed arrows and labeled $T_a$ and $T_b$ in the figure. It can be seen that the electric field distribution  at each scattering-cross-section peak has the number of nodes that would be expected from the order of the corresponding multipole mode. For example, in panel b we have two nodes for the dipole ($m = 1$), though the field pattern is distorted due to the interference with the other modes. It is worth noting that the quality factor increases significantly for modes with higher orders. For instance, the hexapole mode ($m = 3$) with the TM polarization has a full-width-half-maximum FWHM of 10 nm and is located at a center wavelength of 500 nm.

In Fig.~\ref{fig:2dmultipole} (a) and (b), the minima ${T_a}$ and ${T_b}$ are interesting because they are almost scattering-less, a characteristic that is associated with anapoles. With Eq~\ref{eq:14a}, Eq~\ref{eq:13a}, and Eq~\ref{eq:2DsepA} to Eq~\ref{eq:2DsepD}, we can calculate the partial scattering cross-sections of toroidal multipoles. We find that only the dipole terms (i.e. $m=0$ and $m=1$) are non-trivial at ${T_a}$ and ${T_b}$. The partial scattering cross-sections of the expanded terms according to Eq~\ref{eq:2DsepA} to Eq~\ref{eq:2DsepD} are plotted in Fig.~\ref{fig:2dmultipole} (c) and (d). In the figure legends, we use subscript $t$ to denote the toroidal components of the multipole coefficients separated from either the magnetic dipole or the electric dipole. These correspond to the bottom rows of Eqs 5a-5d.  For brevity, we henceforth use the acronyms MTD and ETD to denote the magnetic toroidal dipole and electric toroidal dipole, respectively. In the figure legends, we use the subscripts $m$ and $e$ to denote the dipole coefficients associated with the Cartesian magnetic (i.e. $A_{1,m}$ and $B_{0,m}$) and electric (i.e. $A_{0,e}$ and $B_{1,e}$) dipoles, respectively. For brevity, we also henceforth use the acronyms CMD and CED to denote the Cartesian magnetic and electric dipoles, respectively.

As discussed, from Fig.~\ref{fig:2dmultipole} (a) and (b), it can be seen that coefficients $A_1$ and $B_0$ exhibit dips around $T_a$ and $T_b$, respectively. Fig.~\ref{fig:2dmultipole} (c) and (d) provide insight into this phenomenon. From Fig.~\ref{fig:2dmultipole} (c), it can be seen that the MTD component $A_{1,t}$ has the same magnitude as its CMD counterpart $A_{1,m}$ at $T_a$. This allows destructive interference, provided that the phases differ by a factor $\pi$ (which is indeed the case at $T_a$). Similarly, Fig.~\ref{fig:2dmultipole}(d) reveals that this is also true for the TM case, with the MTD compnent $B_{0,t}$ having the same magnitude as the CMD term $B_{0,m}$ at $T_b$. Destructive interference is again possible because the phases differ by a factor of $\pi$ at $T_b$. We thus conclude that magnetic anapoles occur at wavelengths $T_a$ and $T_b$. 
It can be seen from Fig.~\ref{fig:2dmultipole}(c) that there is another wavelength (other than $T_a$) at which $A_{1,m}$ and $A_{1,t}$ are of equal magnitude. Similarly, there is another wavelength (other than $T_b$) at which $B_{0,m}$ and $B_{0,t}$ match. $A_1$ and $B_0$ do not show zeros at these wavelengths, however, as the phase condition is not met.

We now consider the ETD and CEM components. It can be seen that $A_0$ displays a dip at approx 0.63 $\mu$m. From Fig. 1(c), it can be seen that the magnitude of the ETD component $A_{0,t}$ matches that of the CED component $A_{0,e}$. Destructive interference occurs because the phase condition is met. We also note that while $A_{0,t}$ and $A_{0,e}$ have appreciable magnitudes at shorter wavelengths, $A_0$ is relatively small. This is because the phases of $A_{0,t}$ and $A_{0,e}$ in this spectral range are such that destructive interference occurs. From Fig.~\ref{fig:2dmultipole}(b), it can be seen that $B_1$ is small for wavelengths shorter than 0.6 $\mu$m. This is due to destructive interference between $B_{1,e}$ and $B_{1,t}$. The terms have comparable magnitudes and appropriate phase differences in this wavelength range. It can also be seen that $B_{1,e}$ and $B_{1,t}$ intersect at a wavelength of approx. 0.76 $\mu$m. $B_1$ is not zero at that wavelength however, due to the fact that the phase condition is not met.

\begin{figure}
	\centering
	\fbox{\includegraphics[scale=0.5]{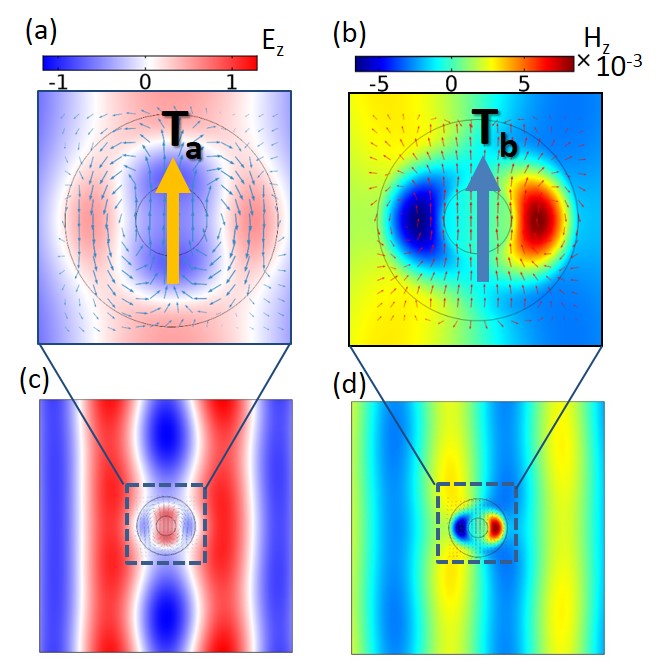}}
	\caption{ (a). Instantaneous electric (E\textsubscript{z}, colour map) and magnetic (H\textsubscript{x} and H\textsubscript{y}, quiver plot) fields around nanotube for TE mode. It can be seen that two loops are formed in the magnetic field pattern. Fields are calculated for illumination wavelength, denoted by $T_a$, in Fig. 1b. Direction of toroidal moment is indicated in this panel by vector also denoted by $\mathbf{T_a}$.(b). Instantaneous magnetic (H\textsubscript{z}, colour map) and electric (E\textsubscript{x} and E\textsubscript{y}, quiver plot) fields around nanotube for TM mode. Calculation is performed for illumination wavelength, denoted by $T_b$, in Fig. 1a. Vector also denoted by $\mathbf{T_b}$ in this panel gives direction of toroidal moment.}
	\label{fig:2dmultipolenf}
\end{figure}

Although the destructive interference of the far-field radiation from the dipole pairs (i.e. (CED, ETD) and (CMD, MTD)) leads to the scattering dips at $T_a$ and $T_b$, there are strong field perturbations locally. These can be seen from the field profiles we present as Fig.~\ref{fig:2dmultipolenf}. The distribution of electric/magnetic field strengths in Fig.~\ref{fig:2dmultipolenf} reveals an intriguing phenomenon. The roles of the magnetic field and the electric field are swapped. In the TE case (Fig.~\ref{fig:2dmultipolenf}a), the toroidal moment can be easily identified by the toroidal distribution of the magnetic field vector, whose double curl defines the direction of the toroidal moment (denoted by the boldface symobol  $\mathbf{T_a}$). On the other hand, for the TM case, it is the double curl of the electric field that defines the direction of the toroidal moment ( denoted by the boldface symbol $\mathbf{T_b}$). Although magnetic anapoles form at wavelengths Ta and Tb due to destructive interference between the CMD and MTD components shown in Fig.~\ref{fig:2dmultipole}, we note that for the TE case, at Ta, the strengths of the fields associated with the CED and ETD are around three times higher than for the CMD and MTD. For the TM case, at Tb, the fields associated with the CED and ETD are comparable to those of the CMD and MTD. In both cases, therefore, the field patterns of Fig. 2 have contributions from all components (CED, CMD, ETD and MTD). The field patterns of Fig.~\ref{fig:2dmultipolenf} are thus not purely magnetic or electric anapole.

\begin{figure}
	
	\centering
	\fbox{\includegraphics[scale=0.8]{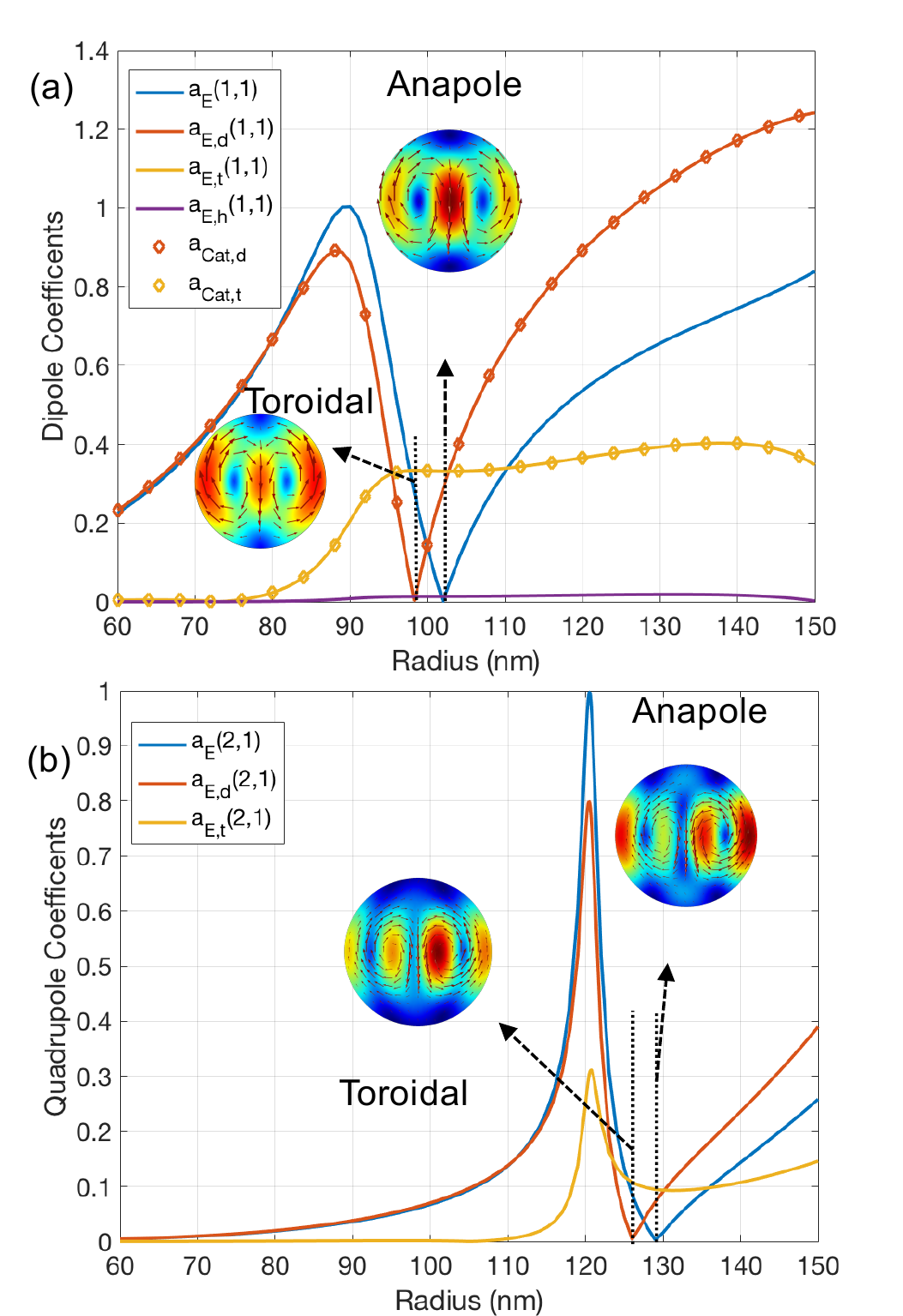}}
	\caption{(a) Spherical multipole coefficients for the dipole moments of order 1 and degree 1. The dipole moments with (solid red line) and without (solid blue line) toroidal moment separation  are calculated with Eq~\ref{eq:16}. The toroidal moment (found using Eq~\ref{eq:22}) is plotted as the solid yellow curve . The curve with red diamond markers is the coefficient converted from Eq~\ref{eq:18} and that with yellow markers is the coefficient converted  from Eq~\ref{eq:19}. The insets are the field patterns of the electric field strength cut at the center of the nano-sphere at two representative wavelengths (the pure toroidal condition and the anapole condition). The solid purple curve plots the next higher order term ($(kr)^5$). (b) The spherical multipole coefficients for the quadrupole moments of order 2 and degree 1.  The quadrupole moments with (solid red line) and without (solid blue line) toroidal quadrupole moment separation  are calculated with Eq~\ref{eq:16}. The toroidal quadrupole moment (found using Eq~\ref{eq:22})is  plotted as the solid yellow curve. The insets are the field patterns of the electric field strength cut at the center of the nano-sphere at the toroidal condition and the anapole condition.}
	\label{fig:3dmultipole}
	
\end{figure}

We now consider a three-dimensional problem. We compare  the results of applying our method to light scattering by a sphere (with anapole excitation) with the results reported by Miroshinichenko et al \cite{miroshnichenko15} using Cartesian expansions. The expansion results from our spherical expansion (Eq~\ref{eq:16} and Eq~\ref{eq:22}) are plotted in Fig.~\ref{fig:3dmultipole}a. They are in agreement with the results calculated and plotted in Fig. 2 of Miroshinichenko et al (the curves with diamond markers in Fig.~\ref{fig:3dmultipole}).

To consider this in further detail, we first recall the equations for the Cartesian electric dipole and the toroidal dipole from Miroshinichenko et al \cite{miroshnichenko15}, 
\begin{eqnarray}
\vec p =&& \frac{i}{\omega}\int_{\bf v} \vec {\bf{J} }d {\bf v} \label{eq:18},
\\
\vec t =&& \frac{1}{10}\int_{\bf v}\left[\left(\vec r \cdotp \vec {\bf{J} }\right)\vec r - 2r^2 \vec {\bf{J} } \right] d {\bf v} \label{eq:19}.
\end{eqnarray}

Comparing these equations with Eq~\ref{eq:16} and \ref{eq:22},  we get: 
\begin{subequations}
	\begin{eqnarray}
	p_x = &&-\frac{3 \pi i}{\omega k^2 \eta} \left[a_{E,d}\left(1,1\right)-a_{E,d}\left(1,-1\right)\right] \label{eq:23}, \\
	p_y = &&-\frac{3 \pi }{\omega k^2 \eta} \left[a_{E,d}\left(1,1\right)+a_{E,d}\left(1,-1\right)\right]  \label{eq:24},\\
	p_z = &&\frac{3\sqrt{2}\pi i}{\omega k^2 \eta} a_{E,d}\left(1,0\right) \label{eq:25},
	\end{eqnarray}
\end{subequations}
\begin{subequations}	
	\begin{eqnarray}
	t_x = && - \frac{3\pi}{k^4 \eta}\left[a_{E,t}\left(1,1\right)-a_{E,t}\left(1,-1\right)\right] \label{eq:26},\\
	t_y = &&- \frac{3\pi}{k^4 \eta i} \left[a_{E,t}\left(1,1\right)+a_{E,t}\left(1,-1\right)\right]  \label{eq:27},\\
	t_z = &&   \frac{3\sqrt{2}\pi}{k^4 \eta} a_{E,t}\left(1,0\right) \label{eq:29}.
	\end{eqnarray}
\end{subequations}

The detailed derivation can be found in Section 4 of the Appendix. Note that in Eq~\ref{eq:16} and \ref{eq:22}, the spherical Bessel function $j_l(kr)$ is expanded only up to order $(kr)^3$, but this fully accounts for the toroidal moments. We have plotted the next higher order moment having the order of $(kr)^5$ in Fig.~\ref{fig:3dmultipole}a as the solid purple curve. We denote it as $a_{E,h}(1,1)$. Although this coefficient is quite small compared to $a_{E,t}(1,1)$ and $a_{E,d}(1,1)$, its presence is ultimately the reason why the anapole condition does not coincide with the perfect cancellation of $a_{E,t}(1,1)$ and $a_{E,d}(1,1)$. This can been from careful examination of Fig.~\ref{fig:3dmultipole}, i.e. the perfect cancellation occurs at slightly longer wavelength than the wavelength at which  $a_{E}(1,1)$ takes a value of zero. The same shift can also be observed from Fig. 2 of the work by Miroshinichenko et al \cite{miroshnichenko15}. The field distribution of $a_{E,h}(1,1)$, however, cannot be identified due to the small strength in comparison to the lower order poles. Nevertheless, we point out that, for larger particles, retaining additional expansion terms might reveal the existence of moments other than the toroidal moment that have even more complicated current configurations but the same radiation pattern.

Next, we show that the Eq~\ref{eq:16} and the Eq~\ref{eq:22} can be used for obtaining any higher order toroidal multipole. We set $l$ to 2 and $m$ to 1 and we solve for electrical quadrupoles.  We plot the calculated quadrupole coefficients in Fig.~\ref{fig:3dmultipole}b. Surprisingly, quadrupolar toroid and anapole show up for a sphere with radius on ~ 30 nm larger than for which the dipole toroid and anapole occur. From the field patterns, we can clearly see the vortices of electric field, similar to the toroidal dipole shown in Fig.~\ref{fig:3dmultipole}a.  Furthermore, there are additional nodes in the field variation, which is due to the quadrupolar nature of the moments. 

Similarly, the toroidal moments of any arbitrary order $l$ and degree $m$ can be found, using the same set of equations (Eq~\ref{eq:16} and Eq~\ref{eq:22} ). The results are trivial in the region of interest studied here thus we do not present them here.  

We note that the magnetic-type anapole was discussed recently by Luk'yanchuk et al.\cite{RN183}.  This anapole was uncovered by comparing the zeros of the Mie coefficients with those of the Cartesian magnetic dipole and the Cartesian magnetic-type toroidal dipole. As shown in Section 5 of the Appendix, the direct expansion of the spherical Bessel function used in our work can also be applied to the spherical magnetic multipole terms, giving the counterpart of the toroidal moment from the magnetic multipole \cite{dubovik90,radescu02,RN184}. We find this term for the dipole can be defined by: 
\begin{equation}
\vec{t}_m = \frac{k^2}{20} \int_{\bf v} \vec{r} \times\vec{r} \times \left(\vec{r} \times \vec {\bf J}\right) d {\bf v}= \frac{k^2}{20} \int_{\bf v} r^2 \vec{r} \times \vec {\bf J} d {\bf v} \label{eq:28}.
\end{equation}

This is exactly the mean square radius of magnetic dipole defined in the literature.\cite{dubovik90,radescu02,RN184} This term has the same radiation pattern as the Cartesian magnetic dipole, while it is composed of toroidal magnetic current distribution defined by the double cross-product of the magnetic moment with position vector $\vec{r}$ (see the middle expression of Eq \ref{eq:28}). It has a dependence of position in the integration of order $(kr)^3$. Note that we have used $kr$ instead of $r$, which is used in most references. The introduction of wavenumber $k$ ensures that the dipole moments can be substituted into formulae (e.g. scattering cross-sections) developed for electric and magnetic dipoles without any scaling factors. In Table 1, we summarize the four multipoles discussed in this paper.  

\section{Conclusions}

In conclusion, we demonstrated a simple but general method to separate higher order terms from curvilinear coordinate systems, including the classical Mie expansion (3D) and the cylindrical expansion (2D). By expanding each electric multipole term with respect to $kr$, we found that the leading term is the Cartesian electric multipole, while the next term is the toroidal multipole. Our expansion approach reveals the existence of terms of order even higher ($>(kr)^3$) than toroidal multipoles. There might be situations (e.g. larger particles) for which these higher order terms provide valuable physical insights. In addition, we discovered a means to readily separate toroidal terms from magnetic multipole expansions. This represents a counterpart to the separation of toroidal terms from electric multipole expansions. 

\begin{table*}
	\caption{A summary of the four types of multipoles discussed in this paper.}
	\begin{tabular}{cccc}
		\hline
		Category & Order of $kr$ & Dipole Expression & Multipole Expression  \\
		\hline
		Cartesian Electric Multipole & $l$ & $\vec{d} = \frac{i}{k}\int_{\bf v} \vec{\bf J}d {\bf v}$ & Eq.~\ref{eq:16}\\
		Cartesian Magnetic Multipole & $l$+1 & $\vec{m} = \frac{1}{2}\int_{\bf v} \vec{r}\times\vec{\bf J}d {\bf v}$ & Eq.~A.46\\
		Electric  Toroidal Multipole & $l$+2 & $\vec{t}_e = \frac{k^2}{10}\int_{\bf v} \left[\left(\vec{r}\cdotp \vec{\bf J}\right)\vec{r}-2r^2\vec{\bf J}\right]d {\bf v}$ & Eq.~\ref{eq:22}\\
		Magnetic  Toroidal Multipole & $l$+3 & $\vec{t}_m = -\frac{k^2}{20}\int_{\bf v} r^2 \vec{r}\times \vec{\bf J}d {\bf v}$ & Eq.~A.47\\
		\hline
	\end{tabular}
	\label{tab:summaryofmultipoles}
\end{table*}

\bibliography{toroidal}

\begin{thebibliography}{34}%
\makeatletter
\providecommand \@ifxundefined [1]{%
 \@ifx{#1\undefined}
}%
\providecommand \@ifnum [1]{%
 \ifnum #1\expandafter \@firstoftwo
 \else \expandafter \@secondoftwo
 \fi
}%
\providecommand \@ifx [1]{%
 \ifx #1\expandafter \@firstoftwo
 \else \expandafter \@secondoftwo
 \fi
}%
\providecommand \natexlab [1]{#1}%
\providecommand \enquote  [1]{``#1''}%
\providecommand \bibnamefont  [1]{#1}%
\providecommand \bibfnamefont [1]{#1}%
\providecommand \citenamefont [1]{#1}%
\providecommand \href@noop [0]{\@secondoftwo}%
\providecommand \href [0]{\begingroup \@sanitize@url \@href}%
\providecommand \@href[1]{\@@startlink{#1}\@@href}%
\providecommand \@@href[1]{\endgroup#1\@@endlink}%
\providecommand \@sanitize@url [0]{\catcode `\\12\catcode `\$12\catcode
  `\&12\catcode `\#12\catcode `\^12\catcode `\_12\catcode `\%12\relax}%
\providecommand \@@startlink[1]{}%
\providecommand \@@endlink[0]{}%
\providecommand \url  [0]{\begingroup\@sanitize@url \@url }%
\providecommand \@url [1]{\endgroup\@href {#1}{\urlprefix }}%
\providecommand \urlprefix  [0]{URL }%
\providecommand \Eprint [0]{\href }%
\providecommand \doibase [0]{http://dx.doi.org/}%
\providecommand \selectlanguage [0]{\@gobble}%
\providecommand \bibinfo  [0]{\@secondoftwo}%
\providecommand \bibfield  [0]{\@secondoftwo}%
\providecommand \translation [1]{[#1]}%
\providecommand \BibitemOpen [0]{}%
\providecommand \bibitemStop [0]{}%
\providecommand \bibitemNoStop [0]{.\EOS\space}%
\providecommand \EOS [0]{\spacefactor3000\relax}%
\providecommand \BibitemShut  [1]{\csname bibitem#1\endcsname}%
\let\auto@bib@innerbib\@empty
\bibitem [{\citenamefont {Zel'Dovich}(1958)}]{zeldovich58}%
  \BibitemOpen
  \bibfield  {author} {\bibinfo {author} {\bibfnamefont {I.~B.}\ \bibnamefont
  {Zel'Dovich}},\ }\href@noop {} {\bibfield  {journal} {\bibinfo  {journal}
  {Sov. J. Exp. Theo. Phys.}\ }\textbf {\bibinfo {volume} {6}},\ \bibinfo
  {pages} {1184} (\bibinfo {year} {1958})}\BibitemShut {NoStop}%
\bibitem [{\citenamefont {Wood}\ \emph {et~al.}(1997)\citenamefont {Wood},
  \citenamefont {Bennett}, \citenamefont {Cho}, \citenamefont {Masterson},
  \citenamefont {Roberts}, \citenamefont {Tanner},\ and\ \citenamefont
  {Wieman}}]{RN108}%
  \BibitemOpen
  \bibfield  {author} {\bibinfo {author} {\bibfnamefont {C.}~\bibnamefont
  {Wood}}, \bibinfo {author} {\bibfnamefont {S.}~\bibnamefont {Bennett}},
  \bibinfo {author} {\bibfnamefont {D.}~\bibnamefont {Cho}}, \bibinfo {author}
  {\bibfnamefont {B.}~\bibnamefont {Masterson}}, \bibinfo {author}
  {\bibfnamefont {J.}~\bibnamefont {Roberts}}, \bibinfo {author} {\bibfnamefont
  {C.}~\bibnamefont {Tanner}}, \ and\ \bibinfo {author} {\bibfnamefont {C.~E.}\
  \bibnamefont {Wieman}},\ }\href@noop {} {\bibfield  {journal} {\bibinfo
  {journal} {Science}\ }\textbf {\bibinfo {volume} {275}},\ \bibinfo {pages}
  {1759} (\bibinfo {year} {1997})}\BibitemShut {NoStop}%
\bibitem [{\citenamefont {Van~Aken}\ \emph {et~al.}(2007)\citenamefont
  {Van~Aken}, \citenamefont {Rivera}, \citenamefont {Schmid},\ and\
  \citenamefont {Fiebig}}]{RN106}%
  \BibitemOpen
  \bibfield  {author} {\bibinfo {author} {\bibfnamefont {B.~B.}\ \bibnamefont
  {Van~Aken}}, \bibinfo {author} {\bibfnamefont {J.-P.}\ \bibnamefont
  {Rivera}}, \bibinfo {author} {\bibfnamefont {H.}~\bibnamefont {Schmid}}, \
  and\ \bibinfo {author} {\bibfnamefont {M.}~\bibnamefont {Fiebig}},\
  }\href@noop {} {\bibfield  {journal} {\bibinfo  {journal} {Nature}\ }\textbf
  {\bibinfo {volume} {449}},\ \bibinfo {pages} {702} (\bibinfo {year}
  {2007})}\BibitemShut {NoStop}%
\bibitem [{\citenamefont {Naumov}\ \emph {et~al.}(2004)\citenamefont {Naumov},
  \citenamefont {Bellaiche},\ and\ \citenamefont {Fu}}]{RN107}%
  \BibitemOpen
  \bibfield  {author} {\bibinfo {author} {\bibfnamefont {I.~I.}\ \bibnamefont
  {Naumov}}, \bibinfo {author} {\bibfnamefont {L.}~\bibnamefont {Bellaiche}}, \
  and\ \bibinfo {author} {\bibfnamefont {H.}~\bibnamefont {Fu}},\ }\href@noop
  {} {\bibfield  {journal} {\bibinfo  {journal} {Nature}\ }\textbf {\bibinfo
  {volume} {432}},\ \bibinfo {pages} {737} (\bibinfo {year}
  {2004})}\BibitemShut {NoStop}%
\bibitem [{\citenamefont {Ho}\ and\ \citenamefont {Scherrer}(2013)}]{RN63}%
  \BibitemOpen
  \bibfield  {author} {\bibinfo {author} {\bibfnamefont {C.~M.}\ \bibnamefont
  {Ho}}\ and\ \bibinfo {author} {\bibfnamefont {R.~J.}\ \bibnamefont
  {Scherrer}},\ }\href@noop {} {\bibfield  {journal} {\bibinfo  {journal}
  {Phys. Lett. B}\ }\textbf {\bibinfo {volume} {722}},\ \bibinfo {pages} {341}
  (\bibinfo {year} {2013})}\BibitemShut {NoStop}%
\bibitem [{\citenamefont {Scagnoli}\ \emph {et~al.}(2011)\citenamefont
  {Scagnoli}, \citenamefont {Staub}, \citenamefont {Bodenthin}, \citenamefont
  {De~Souza}, \citenamefont {García-Fernández}, \citenamefont
  {Garganourakis}, \citenamefont {Boothroyd}, \citenamefont {Prabhakaran},\
  and\ \citenamefont {Lovesey}}]{RN65}%
  \BibitemOpen
  \bibfield  {author} {\bibinfo {author} {\bibfnamefont {V.}~\bibnamefont
  {Scagnoli}}, \bibinfo {author} {\bibfnamefont {U.}~\bibnamefont {Staub}},
  \bibinfo {author} {\bibfnamefont {Y.}~\bibnamefont {Bodenthin}}, \bibinfo
  {author} {\bibfnamefont {R.}~\bibnamefont {De~Souza}}, \bibinfo {author}
  {\bibfnamefont {M.}~\bibnamefont {García-Fernández}}, \bibinfo {author}
  {\bibfnamefont {M.}~\bibnamefont {Garganourakis}}, \bibinfo {author}
  {\bibfnamefont {A.}~\bibnamefont {Boothroyd}}, \bibinfo {author}
  {\bibfnamefont {D.}~\bibnamefont {Prabhakaran}}, \ and\ \bibinfo {author}
  {\bibfnamefont {S.}~\bibnamefont {Lovesey}},\ }\href@noop {} {\bibfield
  {journal} {\bibinfo  {journal} {Science}\ }\textbf {\bibinfo {volume}
  {332}},\ \bibinfo {pages} {696} (\bibinfo {year} {2011})}\BibitemShut
  {NoStop}%
\bibitem [{\citenamefont {Kaelberer}\ \emph {et~al.}(2010)\citenamefont
  {Kaelberer}, \citenamefont {Fedotov}, \citenamefont {Papasimakis},
  \citenamefont {Tsai},\ and\ \citenamefont {Zheludev}}]{kaelberer10}%
  \BibitemOpen
  \bibfield  {author} {\bibinfo {author} {\bibfnamefont {T.}~\bibnamefont
  {Kaelberer}}, \bibinfo {author} {\bibfnamefont {V.}~\bibnamefont {Fedotov}},
  \bibinfo {author} {\bibfnamefont {N.}~\bibnamefont {Papasimakis}}, \bibinfo
  {author} {\bibfnamefont {D.}~\bibnamefont {Tsai}}, \ and\ \bibinfo {author}
  {\bibfnamefont {N.}~\bibnamefont {Zheludev}},\ }\href
  {http://science.sciencemag.org.ezp.lib.unimelb.edu.au/content/sci/330/6010/1510.full.pdf}
  {\bibfield  {journal} {\bibinfo  {journal} {Science}\ }\textbf {\bibinfo
  {volume} {330}},\ \bibinfo {pages} {1510} (\bibinfo {year}
  {2010})}\BibitemShut {NoStop}%
\bibitem [{\citenamefont {Miroshnichenko}\ \emph {et~al.}(2015)\citenamefont
  {Miroshnichenko}, \citenamefont {Evlyukhin}, \citenamefont {Yu},
  \citenamefont {Bakker}, \citenamefont {Chipouline}, \citenamefont
  {Kuznetsov}, \citenamefont {Luk’yanchuk}, \citenamefont {Chichkov},\ and\
  \citenamefont {Kivshar}}]{miroshnichenko15}%
  \BibitemOpen
  \bibfield  {author} {\bibinfo {author} {\bibfnamefont {A.~E.}\ \bibnamefont
  {Miroshnichenko}}, \bibinfo {author} {\bibfnamefont {A.~B.}\ \bibnamefont
  {Evlyukhin}}, \bibinfo {author} {\bibfnamefont {Y.~F.}\ \bibnamefont {Yu}},
  \bibinfo {author} {\bibfnamefont {R.~M.}\ \bibnamefont {Bakker}}, \bibinfo
  {author} {\bibfnamefont {A.}~\bibnamefont {Chipouline}}, \bibinfo {author}
  {\bibfnamefont {A.~I.}\ \bibnamefont {Kuznetsov}}, \bibinfo {author}
  {\bibfnamefont {B.}~\bibnamefont {Luk’yanchuk}}, \bibinfo {author}
  {\bibfnamefont {B.~N.}\ \bibnamefont {Chichkov}}, \ and\ \bibinfo {author}
  {\bibfnamefont {Y.~S.}\ \bibnamefont {Kivshar}},\ }\href@noop {} {\bibfield
  {journal} {\bibinfo  {journal} {Nature Commun.}\ }\textbf {\bibinfo {volume}
  {6}} (\bibinfo {year} {2015})}\BibitemShut {NoStop}%
\bibitem [{\citenamefont {Papasimakis}\ \emph {et~al.}(2016)\citenamefont
  {Papasimakis}, \citenamefont {Fedotov}, \citenamefont {Savinov},
  \citenamefont {Raybould},\ and\ \citenamefont {Zheludev}}]{papsimakis16}%
  \BibitemOpen
  \bibfield  {author} {\bibinfo {author} {\bibfnamefont {N.}~\bibnamefont
  {Papasimakis}}, \bibinfo {author} {\bibfnamefont {V.}~\bibnamefont
  {Fedotov}}, \bibinfo {author} {\bibfnamefont {V.}~\bibnamefont {Savinov}},
  \bibinfo {author} {\bibfnamefont {T.}~\bibnamefont {Raybould}}, \ and\
  \bibinfo {author} {\bibfnamefont {N.}~\bibnamefont {Zheludev}},\ }\href@noop
  {} {\bibfield  {journal} {\bibinfo  {journal} {Nature Mater.}\ }\textbf
  {\bibinfo {volume} {15}},\ \bibinfo {pages} {263} (\bibinfo {year}
  {2016})}\BibitemShut {NoStop}%
\bibitem [{\citenamefont {Fernandez-Corbaton}\ \emph
  {et~al.}(2015{\natexlab{a}})\citenamefont {Fernandez-Corbaton}, \citenamefont
  {Nanz},\ and\ \citenamefont {Rockstuhl}}]{fernandez15}%
  \BibitemOpen
  \bibfield  {author} {\bibinfo {author} {\bibfnamefont {I.}~\bibnamefont
  {Fernandez-Corbaton}}, \bibinfo {author} {\bibfnamefont {S.}~\bibnamefont
  {Nanz}}, \ and\ \bibinfo {author} {\bibfnamefont {C.}~\bibnamefont
  {Rockstuhl}},\ }\href@noop {} {\bibfield  {journal} {\bibinfo  {journal}
  {arXiv preprint arXiv:1507.00755}\ } (\bibinfo {year}
  {2015}{\natexlab{a}})}\BibitemShut {NoStop}%
\bibitem [{\citenamefont {Nanz}(2016)}]{naz16}%
  \BibitemOpen
  \bibfield  {author} {\bibinfo {author} {\bibfnamefont {S.}~\bibnamefont
  {Nanz}},\ }\href@noop {} {\emph {\bibinfo {title} {Toroidal Multipole Moments
  in Classical Electrodynamics: An Analysis of Their Emergence and Physical
  Significance}}}\ (\bibinfo  {publisher} {Springer},\ \bibinfo {year}
  {2016})\BibitemShut {NoStop}%
\bibitem [{\citenamefont {Vrejoiu}(2002)}]{vrejoiu2002}%
  \BibitemOpen
  \bibfield  {author} {\bibinfo {author} {\bibfnamefont {C.}~\bibnamefont
  {Vrejoiu}},\ }\href
  {http://iopscience.iop.org.ezp.lib.unimelb.edu.au/article/10.1088/0305-4470/35/46/313/pdf}
  {\bibfield  {journal} {\bibinfo  {journal} {J. Phys. A. Math. Gen.}\ }\textbf
  {\bibinfo {volume} {35}},\ \bibinfo {pages} {9911} (\bibinfo {year}
  {2002})}\BibitemShut {NoStop}%
\bibitem [{\citenamefont {Radescu}\ and\ \citenamefont
  {Vaman}(2002)}]{radescu02}%
  \BibitemOpen
  \bibfield  {author} {\bibinfo {author} {\bibfnamefont {E.}~\bibnamefont
  {Radescu}}\ and\ \bibinfo {author} {\bibfnamefont {G.}~\bibnamefont
  {Vaman}},\ }\href@noop {} {\bibfield  {journal} {\bibinfo  {journal} {Phys.
  Rev. E}\ }\textbf {\bibinfo {volume} {65}},\ \bibinfo {pages} {046609}
  (\bibinfo {year} {2002})}\BibitemShut {NoStop}%
\bibitem [{\citenamefont {Dubovik}\ and\ \citenamefont
  {Cheshkov}(1967)}]{RN96}%
  \BibitemOpen
  \bibfield  {author} {\bibinfo {author} {\bibfnamefont {V.~I.}\ \bibnamefont
  {Dubovik}}\ and\ \bibinfo {author} {\bibfnamefont {A.~A.}\ \bibnamefont
  {Cheshkov}},\ }\href@noop {} {\bibfield  {journal} {\bibinfo  {journal}
  {Soviet Physics JETP}\ }\textbf {\bibinfo {volume} {24}} (\bibinfo {year}
  {1967})}\BibitemShut {NoStop}%
\bibitem [{\citenamefont {Grahn}\ \emph {et~al.}(2012)\citenamefont {Grahn},
  \citenamefont {Shevchenko},\ and\ \citenamefont {Kaivola}}]{grahn12}%
  \BibitemOpen
  \bibfield  {author} {\bibinfo {author} {\bibfnamefont {P.}~\bibnamefont
  {Grahn}}, \bibinfo {author} {\bibfnamefont {A.}~\bibnamefont {Shevchenko}}, \
  and\ \bibinfo {author} {\bibfnamefont {M.}~\bibnamefont {Kaivola}},\
  }\href@noop {} {\bibfield  {journal} {\bibinfo  {journal} {New J. Phys.}\
  }\textbf {\bibinfo {volume} {14}},\ \bibinfo {pages} {093033} (\bibinfo
  {year} {2012})}\BibitemShut {NoStop}%
\bibitem [{\citenamefont {Mie}(1908)}]{mie1908}%
  \BibitemOpen
  \bibfield  {author} {\bibinfo {author} {\bibfnamefont {G.}~\bibnamefont
  {Mie}},\ }\href@noop {} {\bibfield  {journal} {\bibinfo  {journal} {Annalen
  der physik}\ }\textbf {\bibinfo {volume} {330}},\ \bibinfo {pages} {377}
  (\bibinfo {year} {1908})}\BibitemShut {NoStop}%
\bibitem [{\citenamefont {Evlyukhin}\ \emph {et~al.}(2016)\citenamefont
  {Evlyukhin}, \citenamefont {Fischer}, \citenamefont {Reinhardt},\ and\
  \citenamefont {Chichkov}}]{RN134}%
  \BibitemOpen
  \bibfield  {author} {\bibinfo {author} {\bibfnamefont {A.~B.}\ \bibnamefont
  {Evlyukhin}}, \bibinfo {author} {\bibfnamefont {T.}~\bibnamefont {Fischer}},
  \bibinfo {author} {\bibfnamefont {C.}~\bibnamefont {Reinhardt}}, \ and\
  \bibinfo {author} {\bibfnamefont {B.~N.}\ \bibnamefont {Chichkov}},\
  }\href@noop {} {\bibfield  {journal} {\bibinfo  {journal} {Phys. Rev. B}\
  }\textbf {\bibinfo {volume} {94}},\ \bibinfo {pages} {205434} (\bibinfo
  {year} {2016})}\BibitemShut {NoStop}%
\bibitem [{\citenamefont {Gorbatsevich}\ and\ \citenamefont
  {Kopaev}(1994)}]{RN138}%
  \BibitemOpen
  \bibfield  {author} {\bibinfo {author} {\bibfnamefont {A.}~\bibnamefont
  {Gorbatsevich}}\ and\ \bibinfo {author} {\bibfnamefont {Y.~V.}\ \bibnamefont
  {Kopaev}},\ }\href@noop {} {\bibfield  {journal} {\bibinfo  {journal}
  {Ferroelectrics}\ }\textbf {\bibinfo {volume} {161}},\ \bibinfo {pages} {321}
  (\bibinfo {year} {1994})}\BibitemShut {NoStop}%
\bibitem [{\citenamefont {Eerenstein}\ \emph {et~al.}(2006)\citenamefont
  {Eerenstein}, \citenamefont {Mathur},\ and\ \citenamefont {Scott}}]{RN139}%
  \BibitemOpen
  \bibfield  {author} {\bibinfo {author} {\bibfnamefont {W.}~\bibnamefont
  {Eerenstein}}, \bibinfo {author} {\bibfnamefont {N.}~\bibnamefont {Mathur}},
  \ and\ \bibinfo {author} {\bibfnamefont {J.~F.}\ \bibnamefont {Scott}},\
  }\href@noop {} {\bibfield  {journal} {\bibinfo  {journal} {Nature}\ }\textbf
  {\bibinfo {volume} {442}},\ \bibinfo {pages} {759} (\bibinfo {year}
  {2006})}\BibitemShut {NoStop}%
\bibitem [{\citenamefont {Zimmermann}\ \emph {et~al.}(2014)\citenamefont
  {Zimmermann}, \citenamefont {Meier},\ and\ \citenamefont {Fiebig}}]{RN140}%
  \BibitemOpen
  \bibfield  {author} {\bibinfo {author} {\bibfnamefont {A.~S.}\ \bibnamefont
  {Zimmermann}}, \bibinfo {author} {\bibfnamefont {D.}~\bibnamefont {Meier}}, \
  and\ \bibinfo {author} {\bibfnamefont {M.}~\bibnamefont {Fiebig}},\
  }\href@noop {} {\bibfield  {journal} {\bibinfo  {journal} {Nature Commun.}\
  }\textbf {\bibinfo {volume} {5}} (\bibinfo {year} {2014})}\BibitemShut
  {NoStop}%
\bibitem [{\citenamefont {Hayami}\ \emph {et~al.}(2014)\citenamefont {Hayami},
  \citenamefont {Kusunose},\ and\ \citenamefont {Motome}}]{RN141}%
  \BibitemOpen
  \bibfield  {author} {\bibinfo {author} {\bibfnamefont {S.}~\bibnamefont
  {Hayami}}, \bibinfo {author} {\bibfnamefont {H.}~\bibnamefont {Kusunose}}, \
  and\ \bibinfo {author} {\bibfnamefont {Y.}~\bibnamefont {Motome}},\
  }\href@noop {} {\bibfield  {journal} {\bibinfo  {journal} {Phys. Rev. B}\
  }\textbf {\bibinfo {volume} {90}},\ \bibinfo {pages} {024432} (\bibinfo
  {year} {2014})}\BibitemShut {NoStop}%
\bibitem [{\citenamefont {Basharin}\ \emph {et~al.}(2017)\citenamefont
  {Basharin}, \citenamefont {Chuguevsky}, \citenamefont {Volsky}, \citenamefont
  {Kafesaki},\ and\ \citenamefont {Economou}}]{RN142}%
  \BibitemOpen
  \bibfield  {author} {\bibinfo {author} {\bibfnamefont {A.~A.}\ \bibnamefont
  {Basharin}}, \bibinfo {author} {\bibfnamefont {V.}~\bibnamefont
  {Chuguevsky}}, \bibinfo {author} {\bibfnamefont {N.}~\bibnamefont {Volsky}},
  \bibinfo {author} {\bibfnamefont {M.}~\bibnamefont {Kafesaki}}, \ and\
  \bibinfo {author} {\bibfnamefont {E.~N.}\ \bibnamefont {Economou}},\
  }\href@noop {} {\bibfield  {journal} {\bibinfo  {journal} {Phys. Rev. B}\
  }\textbf {\bibinfo {volume} {95}},\ \bibinfo {pages} {035104} (\bibinfo
  {year} {2017})}\BibitemShut {NoStop}%
\bibitem [{\citenamefont {Raybould}\ \emph {et~al.}(2016)\citenamefont
  {Raybould}, \citenamefont {Fedotov}, \citenamefont {Papasimakis},
  \citenamefont {Kuprov}, \citenamefont {Youngs}, \citenamefont {Chen},
  \citenamefont {Tsai},\ and\ \citenamefont {Zheludev}}]{RN135}%
  \BibitemOpen
  \bibfield  {author} {\bibinfo {author} {\bibfnamefont {T.}~\bibnamefont
  {Raybould}}, \bibinfo {author} {\bibfnamefont {V.}~\bibnamefont {Fedotov}},
  \bibinfo {author} {\bibfnamefont {N.}~\bibnamefont {Papasimakis}}, \bibinfo
  {author} {\bibfnamefont {I.}~\bibnamefont {Kuprov}}, \bibinfo {author}
  {\bibfnamefont {I.}~\bibnamefont {Youngs}}, \bibinfo {author} {\bibfnamefont
  {W.}~\bibnamefont {Chen}}, \bibinfo {author} {\bibfnamefont {D.}~\bibnamefont
  {Tsai}}, \ and\ \bibinfo {author} {\bibfnamefont {N.}~\bibnamefont
  {Zheludev}},\ }\href@noop {} {\bibfield  {journal} {\bibinfo  {journal}
  {Phys. Rev. B}\ }\textbf {\bibinfo {volume} {94}},\ \bibinfo {pages} {035119}
  (\bibinfo {year} {2016})}\BibitemShut {NoStop}%
\bibitem [{\citenamefont {Liu}\ \emph {et~al.}(2017)\citenamefont {Liu},
  \citenamefont {Du}, \citenamefont {Cui}, \citenamefont {Li}, \citenamefont
  {Fan}, \citenamefont {Chen}, \citenamefont {Li}, \citenamefont {Li},\ and\
  \citenamefont {Gu}}]{RN137}%
  \BibitemOpen
  \bibfield  {author} {\bibinfo {author} {\bibfnamefont {Z.}~\bibnamefont
  {Liu}}, \bibinfo {author} {\bibfnamefont {S.}~\bibnamefont {Du}}, \bibinfo
  {author} {\bibfnamefont {A.}~\bibnamefont {Cui}}, \bibinfo {author}
  {\bibfnamefont {Z.}~\bibnamefont {Li}}, \bibinfo {author} {\bibfnamefont
  {Y.}~\bibnamefont {Fan}}, \bibinfo {author} {\bibfnamefont {S.}~\bibnamefont
  {Chen}}, \bibinfo {author} {\bibfnamefont {W.}~\bibnamefont {Li}}, \bibinfo
  {author} {\bibfnamefont {J.}~\bibnamefont {Li}}, \ and\ \bibinfo {author}
  {\bibfnamefont {C.}~\bibnamefont {Gu}},\ }\href@noop {} {\bibfield  {journal}
  {\bibinfo  {journal} {Adv. Mater.}\ } (\bibinfo {year} {2017})}\BibitemShut
  {NoStop}%
\bibitem [{\citenamefont {Mirzaei}\ \emph {et~al.}(2015)\citenamefont
  {Mirzaei}, \citenamefont {Miroshnichenko}, \citenamefont {Shadrivov},\ and\
  \citenamefont {Kivshar}}]{mirzaei15}%
  \BibitemOpen
  \bibfield  {author} {\bibinfo {author} {\bibfnamefont {A.}~\bibnamefont
  {Mirzaei}}, \bibinfo {author} {\bibfnamefont {A.~E.}\ \bibnamefont
  {Miroshnichenko}}, \bibinfo {author} {\bibfnamefont {I.~V.}\ \bibnamefont
  {Shadrivov}}, \ and\ \bibinfo {author} {\bibfnamefont {Y.~S.}\ \bibnamefont
  {Kivshar}},\ }\href
  {http://journals.aps.org/prl/abstract/10.1103/PhysRevLett.115.215501}
  {\bibfield  {journal} {\bibinfo  {journal} {Phys. Rev. Lett.}\ }\textbf
  {\bibinfo {volume} {115}},\ \bibinfo {pages} {215501} (\bibinfo {year}
  {2015})}\BibitemShut {NoStop}%
\bibitem [{\citenamefont {Jackson}(1999)}]{jackson99}%
  \BibitemOpen
  \bibfield  {author} {\bibinfo {author} {\bibfnamefont {J.~D.}\ \bibnamefont
  {Jackson}},\ }\href@noop {} {\emph {\bibinfo {title} {Classical
  electrodynamics}}}\ (\bibinfo  {publisher} {Wiley},\ \bibinfo {year}
  {1999})\BibitemShut {NoStop}%
\bibitem [{\citenamefont {Johnson}\ \emph {et~al.}(2001)\citenamefont
  {Johnson}, \citenamefont {Fan}, \citenamefont {Mekis},\ and\ \citenamefont
  {Joannopoulos}}]{johnson01}%
  \BibitemOpen
  \bibfield  {author} {\bibinfo {author} {\bibfnamefont {S.~G.}\ \bibnamefont
  {Johnson}}, \bibinfo {author} {\bibfnamefont {S.}~\bibnamefont {Fan}},
  \bibinfo {author} {\bibfnamefont {A.}~\bibnamefont {Mekis}}, \ and\ \bibinfo
  {author} {\bibfnamefont {J.}~\bibnamefont {Joannopoulos}},\ }\href@noop {}
  {\bibfield  {journal} {\bibinfo  {journal} {Appl. Phys. Lett.}\ }\textbf
  {\bibinfo {volume} {78}},\ \bibinfo {pages} {3388} (\bibinfo {year}
  {2001})}\BibitemShut {NoStop}%
\bibitem [{\citenamefont {Snyder}(1970)}]{snyder70}%
  \BibitemOpen
  \bibfield  {author} {\bibinfo {author} {\bibfnamefont {A.}~\bibnamefont
  {Snyder}},\ }\href@noop {} {\bibfield  {journal} {\bibinfo  {journal} {IEEE
  Trans. Microw. Theory Tech.}\ }\textbf {\bibinfo {volume} {18}},\ \bibinfo
  {pages} {608} (\bibinfo {year} {1970})}\BibitemShut {NoStop}%
\bibitem [{\citenamefont {Teschl}(2012)}]{RN88}%
  \BibitemOpen
  \bibfield  {author} {\bibinfo {author} {\bibfnamefont {G.}~\bibnamefont
  {Teschl}},\ }\href@noop {} {\emph {\bibinfo {title} {Ordinary differential
  equations and dynamical systems}}},\ Vol.\ \bibinfo {volume} {140}\ (\bibinfo
   {publisher} {American Mathematical Society Providence},\ \bibinfo {year}
  {2012})\BibitemShut {NoStop}%
\bibitem [{\citenamefont {Afanasiev}\ and\ \citenamefont
  {Stepanovsky}(1995)}]{afanasiev95}%
  \BibitemOpen
  \bibfield  {author} {\bibinfo {author} {\bibfnamefont {G.}~\bibnamefont
  {Afanasiev}}\ and\ \bibinfo {author} {\bibfnamefont {Y.~P.}\ \bibnamefont
  {Stepanovsky}},\ }\href@noop {} {\bibfield  {journal} {\bibinfo  {journal}
  {J. Phys. A Math. Gen.}\ }\textbf {\bibinfo {volume} {28}},\ \bibinfo {pages}
  {4565} (\bibinfo {year} {1995})}\BibitemShut {NoStop}%
\bibitem [{\citenamefont {Wang}\ and\ \citenamefont
  {Dal~Negro}(2016)}]{wang16}%
  \BibitemOpen
  \bibfield  {author} {\bibinfo {author} {\bibfnamefont {R.}~\bibnamefont
  {Wang}}\ and\ \bibinfo {author} {\bibfnamefont {L.}~\bibnamefont
  {Dal~Negro}},\ }\href@noop {} {\bibfield  {journal} {\bibinfo  {journal}
  {Opt. Express}\ }\textbf {\bibinfo {volume} {24}},\ \bibinfo {pages} {19048}
  (\bibinfo {year} {2016})}\BibitemShut {NoStop}%
\bibitem [{\citenamefont {Luk'yanchuk}\ \emph {et~al.}(2017)\citenamefont
  {Luk'yanchuk}, \citenamefont {Paniagua-Domínguez}, \citenamefont
  {Kuznetsov}, \citenamefont {Miroshnichenko},\ and\ \citenamefont
  {Kivshar}}]{RN183}%
  \BibitemOpen
  \bibfield  {author} {\bibinfo {author} {\bibfnamefont {B.}~\bibnamefont
  {Luk'yanchuk}}, \bibinfo {author} {\bibfnamefont {R.}~\bibnamefont
  {Paniagua-Domínguez}}, \bibinfo {author} {\bibfnamefont {A.~I.}\
  \bibnamefont {Kuznetsov}}, \bibinfo {author} {\bibfnamefont {A.~E.}\
  \bibnamefont {Miroshnichenko}}, \ and\ \bibinfo {author} {\bibfnamefont
  {Y.~S.}\ \bibnamefont {Kivshar}},\ }\href@noop {} {\bibfield  {journal}
  {\bibinfo  {journal} {Physical Review A}\ }\textbf {\bibinfo {volume} {95}},\
  \bibinfo {pages} {063820} (\bibinfo {year} {2017})}\BibitemShut {NoStop}%
\bibitem [{\citenamefont {Dubovik}\ and\ \citenamefont
  {Tugushev}(1990)}]{dubovik90}%
  \BibitemOpen
  \bibfield  {author} {\bibinfo {author} {\bibfnamefont {V.}~\bibnamefont
  {Dubovik}}\ and\ \bibinfo {author} {\bibfnamefont {V.}~\bibnamefont
  {Tugushev}},\ }\href@noop {} {\bibfield  {journal} {\bibinfo  {journal}
  {Phys. Rep.}\ }\textbf {\bibinfo {volume} {187}},\ \bibinfo {pages} {145}
  (\bibinfo {year} {1990})}\BibitemShut {NoStop}%
\bibitem [{\citenamefont {Fernandez-Corbaton}\ \emph
  {et~al.}(2015{\natexlab{b}})\citenamefont {Fernandez-Corbaton}, \citenamefont
  {Nanz}, \citenamefont {Alaee},\ and\ \citenamefont {Rockstuhl}}]{RN184}%
  \BibitemOpen
  \bibfield  {author} {\bibinfo {author} {\bibfnamefont {I.}~\bibnamefont
  {Fernandez-Corbaton}}, \bibinfo {author} {\bibfnamefont {S.}~\bibnamefont
  {Nanz}}, \bibinfo {author} {\bibfnamefont {R.}~\bibnamefont {Alaee}}, \ and\
  \bibinfo {author} {\bibfnamefont {C.}~\bibnamefont {Rockstuhl}},\ }\href@noop
  {} {\bibfield  {journal} {\bibinfo  {journal} {Optics express}\ }\textbf
  {\bibinfo {volume} {23}},\ \bibinfo {pages} {33044} (\bibinfo {year}
  {2015}{\natexlab{b}})}\BibitemShut {NoStop}%
\end{thebibliography}%

\end{document}